\documentclass[twocolumn]{aa}			
\usepackage{graphicx}
\usepackage{txfonts}
\usepackage{natbib}
\bibpunct{(}{)}{;}{a}{}{,}
\def\gapprox{{_>\atop{^\sim}}}
\def\lapprox{{_<\atop{^\sim}}}

\def\cmmd{\rm {cm^{-3}}}
\def\cmmt{\rm {cm^{-2}}}

\def\s-1{\rm {s^{-1}}}
\def\twco{$^{12}$CO}
\def\thco{$^{13}$CO}

\def\HC3N{HC$_3$N}

\def\kms{\hbox{${\rm km\,s}^{-1}$}}
\def\msun{M$_{\odot}$}
\def\lsun{L$_{\odot}$}

\usepackage{url}

\begin{document}

   \title{Winds of change - a molecular outflow in NGC~1377?}

   \subtitle{The anatomy of an extreme FIR-excess galaxy}

   \author{S. Aalto\inst{1}
          \and
          S. Muller\inst{1}
          \and
          K. Sakamoto\inst{2}
	 \and
          J. S. Gallagher\inst{3}
          \and
	 S. Mart\' in\inst{4}
	 \and
	  F. Costagliola\inst{1}
          }

    \offprints{S. Aalto}

   \institute{Department of Earth and Space Sciences, Chalmers University of Technology, Onsala Observatory,
              SE-439 94 Onsala, Sweden\\
              \email{saalto@chalmers.se}
	\and
	Institute of Astronomy and Astrophysics, Academia Sinica,
   	 P.O. Box 23-141, Taipei 10617, Taiwan
	\and
        Department of Astronomy, University of Wisconsin-Madison, 5534 Sterling, 475 North Charter Street, Madison WI 53706, USA
        \and
	European Southern Observatory, Alonso de C\' ordova 3107, Vitacura, Casilla, 19001, Santiago 19, Chile
	  }

   \date{Received x; accepted y}
 
  \abstract
   {}
{Our goal was to investigate the molecular gas distribution and kinematics in the extreme FIR-excess galaxy NGC~1377
and to address the nature and evolutionary status of the buried source.}
{We use high ($0.''65 \times 0.''52$, ($65 \times 52$ pc)) and low ($4.'' 88 \times 2.'' 93$) resolution SubMillimeter Array
(SMA) observations to image the \twco\ and \thco\ 2--1 line emission.}
{We find bright, complex \twco\ 2--1 line emission in the inner 400~pc of NGC~1377. The \twco\ 2--1 line has wings that 
are tracing a kinematical component which appears perpendicular to the component traced by the line core. Together with
an intriguing X-shape of the integrated intensity and dispersion maps, this suggests that the molecular emission of NGC~1377
consists of a disk-outflow system. Lower limits to the molecular mass and outflow rate are 
$M_{\rm out}$(H$_2$)$>$$1 \times 10^7$ \msun\ and $\dot{M}$$>$8 \msun\ yr$^{-1}$. The age of the proposed outflow
is estimated to 1.4~Myrs, the extent to 200~pc and the outflow speed to $V_{\rm out}$=140 \kms. 
The total molecular mass in the SMA map is estimated to $M_{\rm tot}$(H$_2$)=$1.5 \times 10^8$ \msun\ (on a scale of 400~pc) while in the inner
$r$=29~pc the molecular mass is $M_{\rm core}$(H$_2$) $=1.7 \times 10^7$ \msun\ with a corresponding  H$_2$ column density of
$N$(H$_2$)=$3.4 \times 10^{23}$ $\cmmt$ and an average \twco\ 2--1 brightness temperature of 19~K. 
\thco\ 2--1 emission is found at a factor 10 fainter than \twco\ in the low resolution map while C$^{18}$O 2--1 remains undetected. 
We find weak 1~mm continuum emission of 2.4~mJy with spatial extent less than 400~pc.}
{Observing the molecular properties of the FIR-excess galaxy NGC~1377 allows us to probe the early stages of nuclear activity and the onset of feedback 
in active galaxies. The age of the outflow supports the notion that the current nuclear activity is young - a few Myrs. The outflow may 
be powered by radiation pressure from a compact, dust enshrouded nucleus, but other driving mechanisms are possible.
The buried source may be an AGN or an extremely young (1~Myr) compact starburst. Limitations on size and mass
lead us to favour the AGN scenario,  but further studies are required to settle the issue. In either case, the wind with its implied mass
outflow rate will quench the nuclear power source within a very short time of 5--25 Myrs.  It is
however possible that the gas is unable to escape the galaxy and may eventually fall back onto NGC~1377 again.
}

   \keywords{galaxies: evolution
--- galaxies: individual: NGC~1377
--- galaxies: active
--- galaxies: starburst
--- radio lines: ISM
--- ISM: molecules
}
\titlerunning{CO in NGC~1377}
\maketitle

\section{Introduction}
\label{s:intro}

A small subset of galaxies deviate strongly from the well-known radio-to-FIR correlation through having
excess FIR emission as compared to the radio ($q>3$; $q$=log{[FIR/3.75$\times 10^{12}$~Hz]/$S_{\nu}$(1.4GHz)} \citep{helou85}).
There are several potential interpretations of the excess including very young synchrotron-deficient starbursts
or dust enshrouded Active Galactic Nuclei (AGN).  AGNs obscured by nuclear dust are likely also in the early stages of their evolution where nuclear material
has not yet been dispersed and/or consumed to feed the growth of the black hole.
The FIR-excess galaxies are rare - \citet{roussel03} find that they represent a small fraction (1\%) of an infrared flux-limited sample in the local universe,
such as the IRAS Faint Galaxy Sample.  Despite their scarcity, which likely is an effect of the short time 
spent in the FIR-excess phase, these objects deserve careful study.
Their implied youth provides an ideal setting to better understand the initial conditions and early evolution
of starburst and/or AGN activity, as well as insights into how the infrared and radio emission is regulated in galaxies.

\subsection{The extreme FIR-excess galaxy NGC~1377}

NGC~1377 is a member of the Eridanus galaxy group at an estimated distance of 21~Mpc (1\arcsec=102~pc) and has
a far-infrared luminosity of $L_{\rm FIR}=4.3 \times 10^9$ L$_{\sun}$ \citep{roussel03}.
In stellar light, NGC~1377 has the appearance of a regular lenticular galaxy \citep{rc3}, with a diameter
of 1$'.8$, a large scale inclination of 60$^{\circ}$ and a major axis position angle (PA) of 92$^{\circ}$ 
(as derived from the K-band image).  However, \citet{heisler94} reported the
presence of a faint dust lane along the southern part of the minor axis, perturbing an otherwise featureless
morphology.  NGC~1377 is a Sixty Micron Peaker (SMP) meaning that its IR SED peaks near 60$\mu$m.
SMPs are often found to have peculiar morphologies and are either classified as H~II region like or Sy~2 - where
the Sy~2 are somewhat more common \citep[e.g.][]{heisler94, laureijs}. 

NGC~1377 is {\it the most extreme example known so far of an FIR-excess galaxy} with 
radio synchrotron emission being deficient by at least a factor of 37 with 
respect to normal galaxies \citep{roussel03,roussel06} (corresponding to $q > 3.92$). Interestingly,  H~II regions
were not detected through near-infrared hydrogen recombination lines or thermal radio continuum \citep{roussel03,roussel06}. 
The presence of cold molecular gas in the center of NGC~1377 is evident through bright single-dish \twco\ 1--0 and 2--1 line emission
detected by \citet{roussel03}. Deep mid-infrared silicate absorption features suggest that the nucleus is
enshrouded by large masses of dust \citep[e.g.][]{spoon07} potentially  absorbing all of the ionizing photons.
The extremely high obscuration aggravates the determination of the nature of the nuclear activity.

\citet{roussel06} propose that NGC~1377 is a nascent ($t\lapprox$1~Myr) opaque starburst - the radio
synchrotron deficiency caused by the extreme youth (pre-supernova stage) of the starburst activity where
the young stars are still embedded in their birth-clouds. 
In contrast, \citet{imanishi06} argue, based on the small 3.3 $\mu$m PAH equivalent widths and very red $L$-band continuum,
that NGC~1377 harbours a buried AGN.
Furthermore, \citet{imanishi09} find an HCN/HCO$^+$ $J$=1--0 line ratio
exceeding unity, which they suggest is evidence of an X-ray Dominated Region (XDR) surrounding an AGN.
They explain the lack of radio continuum through suggesting the presence of a large column of intervening material causing
free-free absorption. Note, however, that the HCN/HCO$^+$ line ratio
determination of the  XDR contribution is uncertain and is based on small differences in the ratios
(see e.g. Fig. 5 in \citet{krips08}).  Furthermore, the lack of a thermal radio continuum detection also put limits on
the possibility of nuclear free-free absorption.

To further our understanding of the nature of NGC~1377 we obtained high resolution \twco\ and \thco\ 2--1 data with
the SubMillimeter Array (SMA) in Hawaii. 
As NGC~1377 remains undetected in HI and H$\alpha$ according to reports in the literature\footnote{Gallagher et al (2012 in preparation) obtained
a spectrum with the Southern African Large Telescope (SALT) that shows emission for H$\alpha$ and [N II], and we therefore assume that these
lines, along with the previously measured [S II] double, are weak but present in NGC~1377.} \citep{roussel03,roussel06} 
the molecular lines seem to be an important way to study the gas properties and dynamics. 
Our goal was to search for clues to the extreme FIR-excess
of NGC~1377 through the distribution and kinematics of the molecular gas. We found evidence of a molecular
disk-outflow system and we we use its properties to address the nature of the buried source. \\


In Sect.~\ref{s:obs} we present the observations and results which are discussed in terms of 
the structure and power of a molecular disk-outflow system in Sect.~\ref{s:dyn} and Sect.~ \ref{s:power}.
In Sect.\ref{s:driving} possible power sources of the outflow are presented and in Sect.~\ref{s:comp} we briefly compare
with molecular outflows in the literature. The properties of the nuclear gas are discussed in Sect.~\ref{s:nuclear}  and
evolutionary implications for NGC~1377 are discussed in Sect.~\ref{s:evolution}.


\section{Observations and results}
\label{s:obs}

NGC~1377 was observed with the SubMillimeter Array (SMA) on 2009 July 17th, in the
very-extended configuration (8 antennas), and on 2009 October 12th, in the compact
configuration (7 antennas). The phase center was set at $\alpha$=03:36:39.10 and 
$\delta$=$-$20:54:08.0 (J2000). During both nights, the zenith atmospheric opacity at
225~GHz was near 0.1, resulting in system temperatures in the range 100 -- 200 K
depending on source elevation.

The heterodyne SIS receivers were tuned to the frequency of the \twco\ 2--1
transition at 230.538~GHz in the upper sideband, while the \thco\ 2--1
transition was observed in the lower sideband.
The correlator was configured to provide a spectral resolution of 0.8125~MHz.

The bandpass of the individual antennas was derived from the bright quasars J$1924-292$
in July, and 3C454.3 in October. The primary flux calibration was set on Callisto
and Uranus, respectively. The close-by quasars J0423-013 ($\sim$4~Jy at 1~mm) and
J0334-401 ($\sim$1.3~Jy) were observed regularly for complex gain calibration,
every 15 minutes for observations in the very-extended configuration, and every
25 minutes for the compact configuration. We derived a flux density for J0423-013 of 3.3~Jy in
July and 4.5~Jy in October and for  J0334-401 the derived flux density was 1.2~Jy in July and 1.4~Jy in October.
The quasar J0340-213, closer to NGC~1377 ($1^\circ$ away) but weaker ($\sim$0.4~Jy)
than J0423-013 and J0334-401 was also observed during the very-extended configuration
track for calibration-testing purpose: after the gain solution derived from J0423-013 and
J0334-401 was applied to J0340-213, it was found with no measurable offsets relative to its
known position, and with a consistent flux, making us confident with the calibration of our
very-extended configuration data.

After calibration within the dedicated MIR/IDL SMA reduction package, both visibility
sets were converted into FITS format, and imported in the GILDAS/MAPPING and AIPS packages
for further imaging.


\begin{figure*}
\resizebox{17cm}{!}{\includegraphics[angle=0]{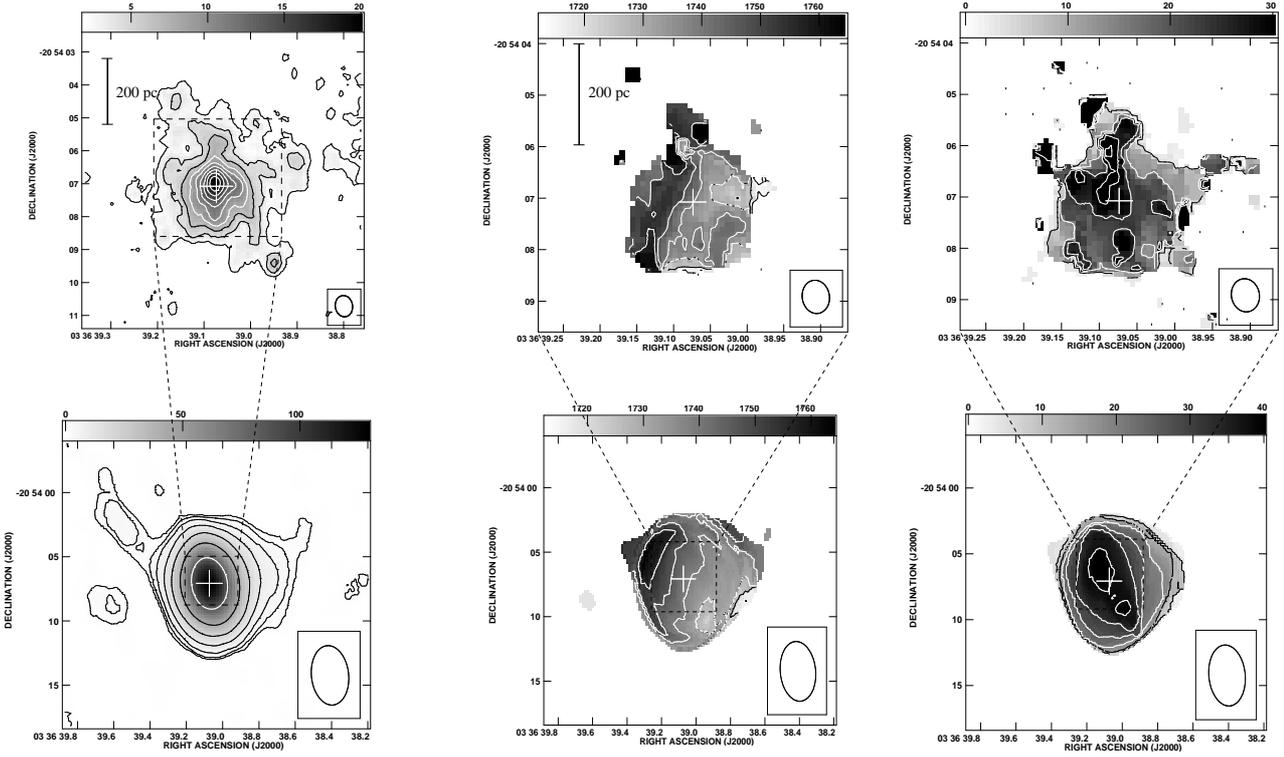}}
\caption{\label{f:mom} {\bf Top panels:} High resolution moment maps of \twco\ 2--1. {\it Left:} Integrated intensity with contour levels 
$2 \times (1,2,3,4,5,6,7,8,9)$ Jy~beam$^{-1}$~\kms\ and gray scale ranging from 2 to 20 Jy~beam$^{-1}$~\kms. The cross marks the position
$\alpha$:03:36:39.073 $\delta$:-20:54:07.08, which is the position of peak $T_{\rm B}$(\twco\ 2--1). {\it Centre:}  Velocity field with
contour levels starting at 1713~\kms\ and then increasing
by steps of 8.6~\kms. The grayscale ranges from 1713 to 1764~\kms. {\it Right:} Dispersion map where contours start at 5~\kms\ and then increase
by steps of 5~\kms. Grayscale ranges from 0 to 30~\kms\  and peak dispersion is 90~\kms. {\bf Lower panels:} Low resolution moment maps of
\twco\ 2--1.{\it Left:}  Integrated intensity with contour levels 
$1.3 \times (1,2,4,8,16,32,64)$ Jy~beam$^{-1}$~\kms\ and gray scale ranging from 0 to 129 Jy~beam$^{-1}$~\kms. 
{\it Centre:} Velocity field with contour levels starting at 1713~\kms\ and then increasing
by steps of 8.6~\kms. The grayscale ranges from 1713 to 1764 \kms. {\it Right:}  Dispersion map where contours start at 6.4~\kms\ and then increase
by steps of 6.4~\kms. Grayscale ranges from 0 to 40~\kms\  and peak dispersion is 65~\kms. Note that the high resolution moment 1 and 2 maps
are on a somewhat different scale compared to the moment 0 map to better show the details of the velocity field and dispersion map.
}
\end{figure*}


\subsection{\twco\ 2--1 emission}

For the \twco\ 2--1 data, sets of visibilities were combined and deconvolved using the
Clark method with uniform weighting. This results in a synthesized beam size of
$0.\arcsec 65 \times 0.\arcsec 52$ with position angle PA=$8^\circ$. We smoothed the data to
a velocity resolution of 5 \kms, yielding a $1\sigma$ rms noise level of 16~mJy~beam$^{-1}$.

Furthermore, data taken with the compact array were also deconvolved with natural weighting resulting 
in a synthesized beam size of $4.\arcsec 88 \times 2. \arcsec 93$ (PA$=176^\circ$)($1\sigma$ rms noise
level of 23~mJy~beam$^{-1}$). This
provides a low resolution map to which we compare the high resolution data. 

\begin{figure*}
\resizebox{15cm}{!}{\includegraphics[angle=0]{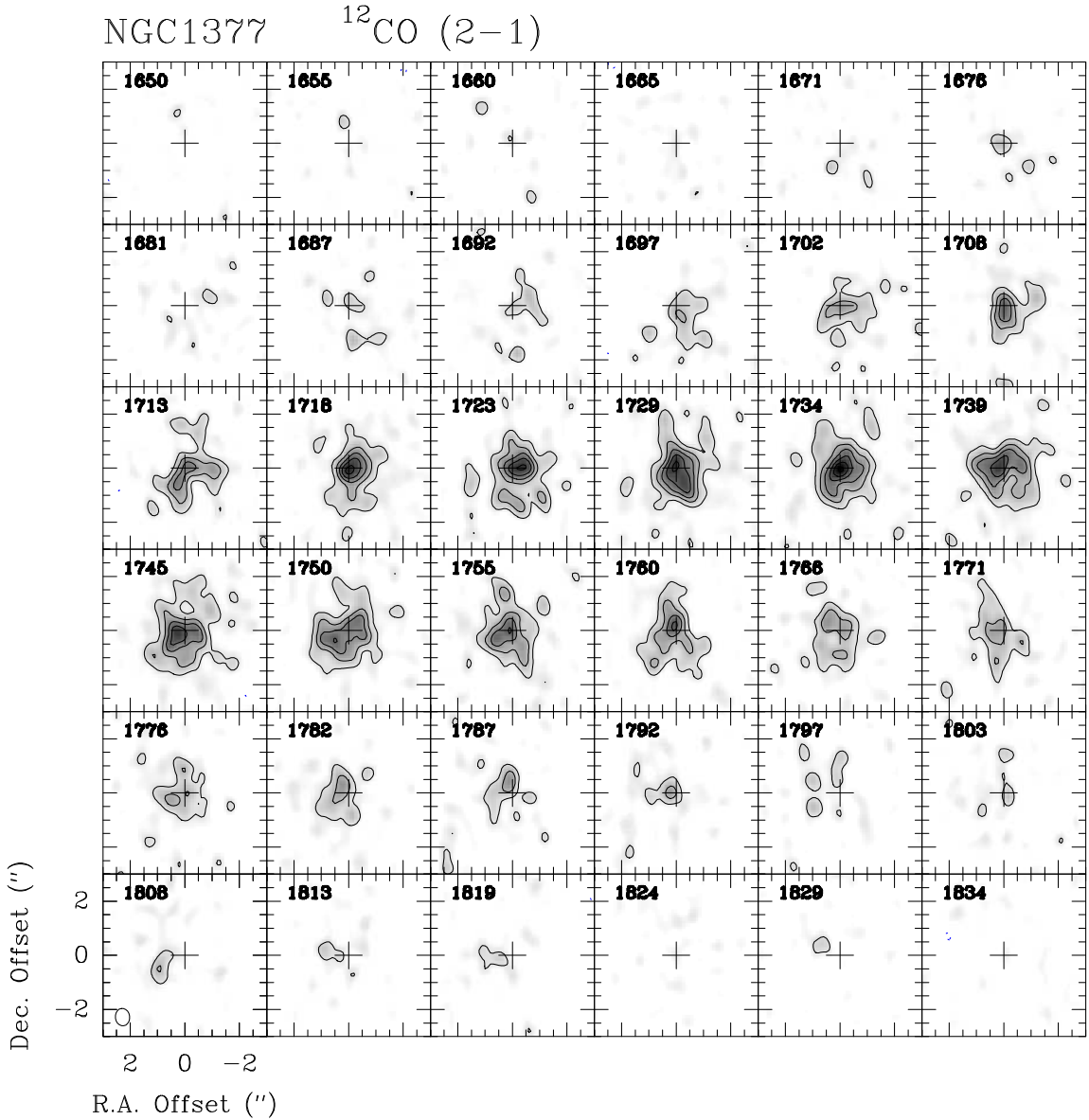}}
\caption{\label{f:chan_co} High resolution channel map of the \twco\ 2--1 emission toward NGC~1377,
as obtained from the combination of the compact and very-extended
configuration data. Contours are drawn every 48 mJy~beam$^{-1}$ (3$\sigma$).
The synthesized beam of $0.''65 \times 0.''52$ is shown in the bottom left corner.}
\end{figure*}

The integrated intensity maps, velocity fields and dispersion maps are presented in Fig.~\ref{f:mom}
and the high resolution channel maps are shown in Fig.~\ref{f:chan_co}.

\subsubsection{The integrated \twco\ 2--1 intensity and molecular mass}

In the high resolution map (Fig.~\ref{f:mom} , top left) the molecular emission is resolved and arises mostly from a region of 1\arcsec\ (100~pc
at $D$=21~Mpc) in radius. (From a  Gaussian fit we find a full width half maximum (FWHM) major axis of $1.\arcsec 76$ and minor
axis $1.\arcsec 4$). Fainter emission extends out to a radius of 2--3\arcsec. The integrated emission structure  is quite complex with various filamentary
features extending out from the center.  A tilted X-like shape appears to emerge from the center. In the low resolution map the emission
is mostly unresolved, but some faint emission is extending to the west and north-east from the central condensation.

The integrated intensities and estimated molecular masses are presented in Tab.~\ref{t:flux}. 
The H$_2$ column density towards the inner $D$=58~pc (average beam diameter) is estimated to $N$(H$_2$)= $3.4 \times 10^{23}$ $\cmmt$
leading to an average nuclear gas surface density of $\Sigma \approx 5 \times 10^3$ \msun pc$^{-2}$.
The average column density for the central 390~pc is $\approx 5 \times 10^{22}$ $\cmmt$. Please see footnote to Tab.~\ref{t:flux} for
\twco\ to H$_2$ mass conversion. Potential errors are further discussed in Sect.~\ref{s:mass_wing}.

We investigated what fraction of the single dish flux was recovered with our interferometric
observations by comparing with our IRAM 30m \twco\ 2--1 observations. We find a \twco\ 2--1 luminosity
in the 9\arcsec\ IRAM beam of $3.5 \times 10^7$~K~\kms~pc$^2$ - to be compared to the SMA total \twco\ luminosity
of $3.3 \times 10^7$~K~\kms~pc$^2$. Within the errors we conclude that all of the  IRAM 30m flux has
been recovered with the interferometer.

\begin{table}
\caption{\label{t:flux} {\bf  \twco\ 2--1 flux densities and molecular masses$^a$}}
\begin{tabular}{ll}
 & \\
\hline
\hline \\ 
Position$^b$ (J2000) & $\alpha$:  03:36:39.076 ($\pm$ 0.\arcsec 2) \\
                           & $\delta$: -20:54:07.01 ($\pm$ 0.\arcsec 1) \\
Peak flux density$^c$ & 270 $\pm$ 16 (mJy\,beam$^{-1}$)\\ 
Flux$^d$ &  \\
\, \, (central beam) & $20 \pm 0.05$ (Jy \kms)\\
\, \, (whole map) & $169 \pm 0.5$ (Jy \kms)\\
\hline \\ 
Molecular mass$^e$ & \\
\, \, (central beam) & $1.7 \times 10^7$ \msun\\
\, \, (whole map) & $1.5 \times 10^8$ \msun\\
\\

\hline \\
\end{tabular} 

{\it a)}\, The high resolution  \twco\ 2--1 data. Listed  errors are 1$\sigma$ rms.

{\it b)}\, The position of the \twco\ 2--1 integrated intensity. The peak
$T_{\rm B}$ is at $\alpha$:03:36:39.073 $\delta$:-20:54:07.08 at $V_c$=1734~\kms. 

{\it c)}\, The Jy to K conversion in the $0.''65 \times 0.52''$ beam
is 1~K=14.5~mJy.

{\it d)}\,  A two-dimensional Gaussian was fitted to the integrated intensity image. 

{\it e)}\,  The \twco\ luminosity $L$(\twco)=$18.5 \times D^2 \times (\theta)^2 \times I$(\twco) K~\kms~pc$^2$
(where $D$ is in Mpc, $\theta$ in \arcsec and $I$(\twco) in K \kms) and M(H$_2$)=3.47$L$(\twco)
for $N$(H$_2$)/$I$(\twco)=$2.5 \times 10^{20}$ $\cmmt$).
The conversion factor has been calibrated for \twco\ 1--0 emission- thus, if the \twco\ is subthermally excited
the H$_2$ mass estimated from the 2--1 line should be corrected upwards. From our IRAM \twco\ 2--1 and 1--0 data
we estimate that the 2--1/1--0 line ratio to 0.7 - and hence we correct out masses upwards by 30\%.

\end{table}

\subsubsection{Kinematics}
\label{s:kinematics}

{\it Velocity field:} \, The high resolution velocity field (Fig.~\ref{f:mom}, top centre) shows complex dynamics dominated by red-shifted gas to the east of the center
and blue-shifted gas to the west. Fitted velocity centroids range from 1690 to 1770~\kms. In the low resolution map, the position angle of
the smoother velocity field is dominated by a PA$\approx$$40^{\circ}$ component and velocity centroids range from 1712 to 1764 \kms.
The velocity centroids were determined through a flux weighted first moment of the spectrum of each pixel therefore assigning one
velocity to a potentially complex spectral structure.  \\

\noindent
{\it Dispersion maps:} \, The high resolution dispersion map (Fig.~\ref{f:mom}, top right)  shows a distinct X-like shape - where the ``x'' itself has a PA of
35$^{\circ}$ -- 40$^{\circ}$. Dispersion (one dimensional, $\sigma_{\rm 1d}$) in the X-structure is typically 25~\kms, but - reaches 44~\kms\ or higher in isolated regions.
The large dispersion is caused by multiple, narrow, spectral features and/or broad wing-like components.
The low resolution dispersion map shows a less complicated pattern where the dispersion is dominated by a bipolar large scale
feature  with peaks of $\sigma_{\rm 1d}$=40~\kms. 
The dispersion was determined through a flux weighted second moment  of the spectrum of each pixel. This corresponds to the one-dimensional velocity
dispersion (i.e. the FWHM line width of the spectrum divided by 2.35 for a Gaussian line profile).
\\

\noindent
{\it Spectra and channel map:} \, The low resolution integrated \twco\ 2--1 spectrum is characterized by narrow emission at the
line center with FWHM line width $\delta V$$\approx$60 \kms\ and by triangular-shaped wings (or plateau)
ranging from 1620~\kms\ to 1860~\kms\ (see Fig.~\ref{f:spec_tco} top panel).  Wings are therefore stretching out to $\pm 120$ \kms\ (projected). 
This line shape is also seen in the SEST single-dish \twco\ 1--0 spectrum by \citet{roussel03}. 
The high resolution channel map reveals significant structure and complexity in the \twco\ 2--1 line emission
per channel. Note that there is no evidence of a bright, broad compact emission from an unresolved nuclear source.
Brightest line emission in terms of peak $T_{\rm B}$ (19~K) occurs at $V$=1734~\kms\ which we adopt as the center velocity. \\

\noindent
{\it Line wings:} \, To identify the spatial origin of the line wings, we integrated the emission corresponding
to velocities 1660 -- 1676 km~s$^{-1}$ (blue wings) and 1813 -- 1829 km~s$^{-1}$ (red wings) ($V_{\rm sys}$=1734 \kms)
in the low resolution \twco\ 2--1 data. The resulting map
is presented as the top left panel of Fig.~\ref{f:blue-red}. The blue and red wings are
spatially separated by $\approx$2\arcsec and with a PA of 45$^{\circ}$.
In the lower left  panel of Fig.~\ref{f:blue-red} we present an image from the high resolution data. Here we have integrated
the emission corresponding to the central channels: blue:1718-1729 \kms\ and red: 1739-1750 \kms. In this figure, we see
a velocity shift along an axis perpendicular to the one seen for the line wings of the low resolution data. 

The relative positional uncertainty is on the order of
(beam size)/(s/n ratio). For the low resolution wing map this is less than 0.\arcsec 7 and for the high res map it is 0.\arcsec 05 - 0.\arcsec 1 .
Therefore the two orthogonal velocity gradients are observationally robust.

\subsection{\thco\ and C$^{18}$O 2--1 emission}

\begin{figure}
\resizebox{8cm}{!}{\includegraphics[angle=0]{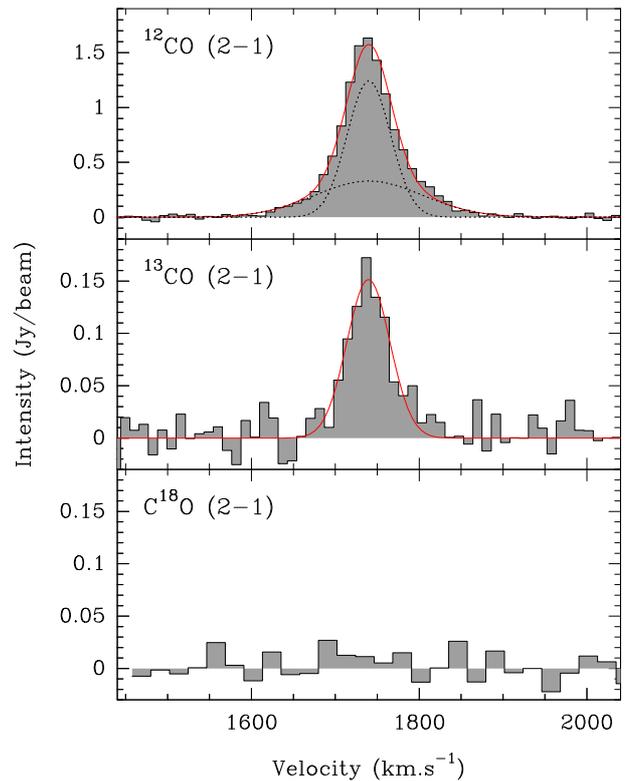}}
\caption{\label{f:spec_tco} Spectra of {\it Top panel:} \twco\ 2--1, {\it Middle panel:}  \thco\ 2--1
and {\it Lower panel:} C$^{18}$O 2--1. All three spectra are taken at position $\alpha$=03:36:39.07
$\delta$-20:54:07.00 in the low resolution data. \thco\ is clearly detected and a factor of 9--10
fainter than \twco. C$^{18}$O is not detected and the \twco/C$^{18}$O 2--1 peak line intensity ratio
is $>$50. The fits to the \twco\ and \thco\ spectra are discussed in Sect.~\ref{s:mass_wing}
}
\end{figure}

For \thco\  and C$^{18}$O 2--1, as the signal-to-noise ratio is lower, we used only data from the compact
configuration, deconvolved using natural weighting and smoothed to a velocity resolution of
11 km~s$^{-1}$. The resulting beam size is $4.\arcsec 88 \times 2.\arcsec 93$ (P.A. $= 176^\circ$) and
the $1\sigma$ rms noise level is 13~mJy~beam$^{-1}$. The low resolution \thco\ and C$^{18}$O spectra
are presented in Fig.~\ref{f:spec_tco}. The \thco\ is clearly detected 
and a factor of 9--10 fainter than \twco. C$^{18}$O is not detected and the \twco/C$^{18}$O 2--1 peak line
intensity ratio is $>$50.  The integrated intensity and velocity field maps are presented
in Fig.~\ref{f:13co}. The \thco\ peak integrated emission is offset by $0.\arcsec 9$ (90~pc) to the south-east from
the peak integrated \twco\ emission. Furthermore,  the \thco\ velocity field appears more consistent with that of the
high-resolution \twco\ data.

\subsection{Continuum emission}

We detect (4$\sigma$) weak 225~GHz continuum in NGC~1377 of 2.4 ($\pm 0.45$) mJy~beam$^{-1}$ in the compact
array at the position of the \twco\ 2--1 line emission. The source is unresolved in the $4.\arcsec 88 \times 2. \arcsec 93$ beam.


\begin{figure}
\resizebox{8cm}{!}{\includegraphics[angle=0]{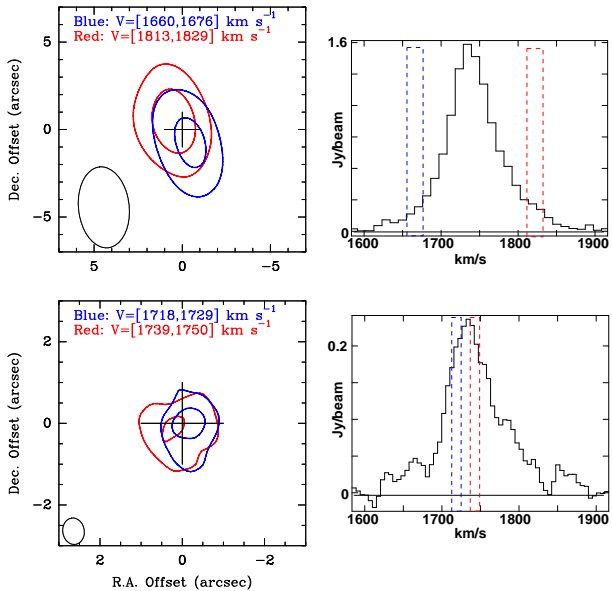}}
\caption{\label{f:blue-red} {\bf Left: } {\it Top panel:}  Blue (dark) ($V$= 1660 -- 1676 km~s$^{-1}$) and red (light) ($V$= 1813 -- 1829 km~s$^{-1}$)
showing the location of the wings of the global \twco\ spectrum. This is done with the low resolution data for maximum 
signal to noise. {\it Lower panel:}  Blue (dark) ($V$= 1718 -- 1729 km~s$^{-1}$)  and red (light) ($V$= 1739 -- 1750 km~s$^{-1}$) part of the central
part of the spectrum of the high resolution data.
The velocities are selected to not overlap with the line wings. Contours are given every $3\sigma$.
Two velocity components - perpendicular to each other - can be discerned. {\bf Right:} {\it Top panel:}  Low resolution \twco\  2--1 spectrum where the
spectral regions in the line wings used to make the figure to the left are indicated. {\it Lower panel:}  High resolution \twco\ 2--1 spectrum where the
spectral regions in the line center used to make the figure to the left are indicated.  }
\end{figure}

\begin{figure}
\resizebox{9cm}{!}{\includegraphics[angle=0]{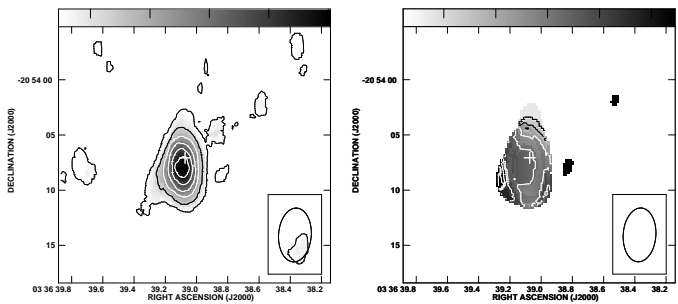}}
\caption{\label{f:13co} Low resolution moment maps of
\thco\ 2--1. {\it Left panel:} Integrated intensity with contour levels 
$0.84 \times (1,3,5,7,9,11)$ Jy~beam$^{-1}$~\kms\ and gray scale ranging from 1 to 10 Jy~beam$^{-1}$~\kms. 
{\it Right panel:}  Velocity field with contour levels starting at 1717~\kms\ and then increasing
by steps of 8.6~\kms. The grayscale ranges from 1713 to 1764 \kms. 
}
\end{figure}

\begin{figure}
\resizebox{8cm}{!}{\includegraphics[angle=0]{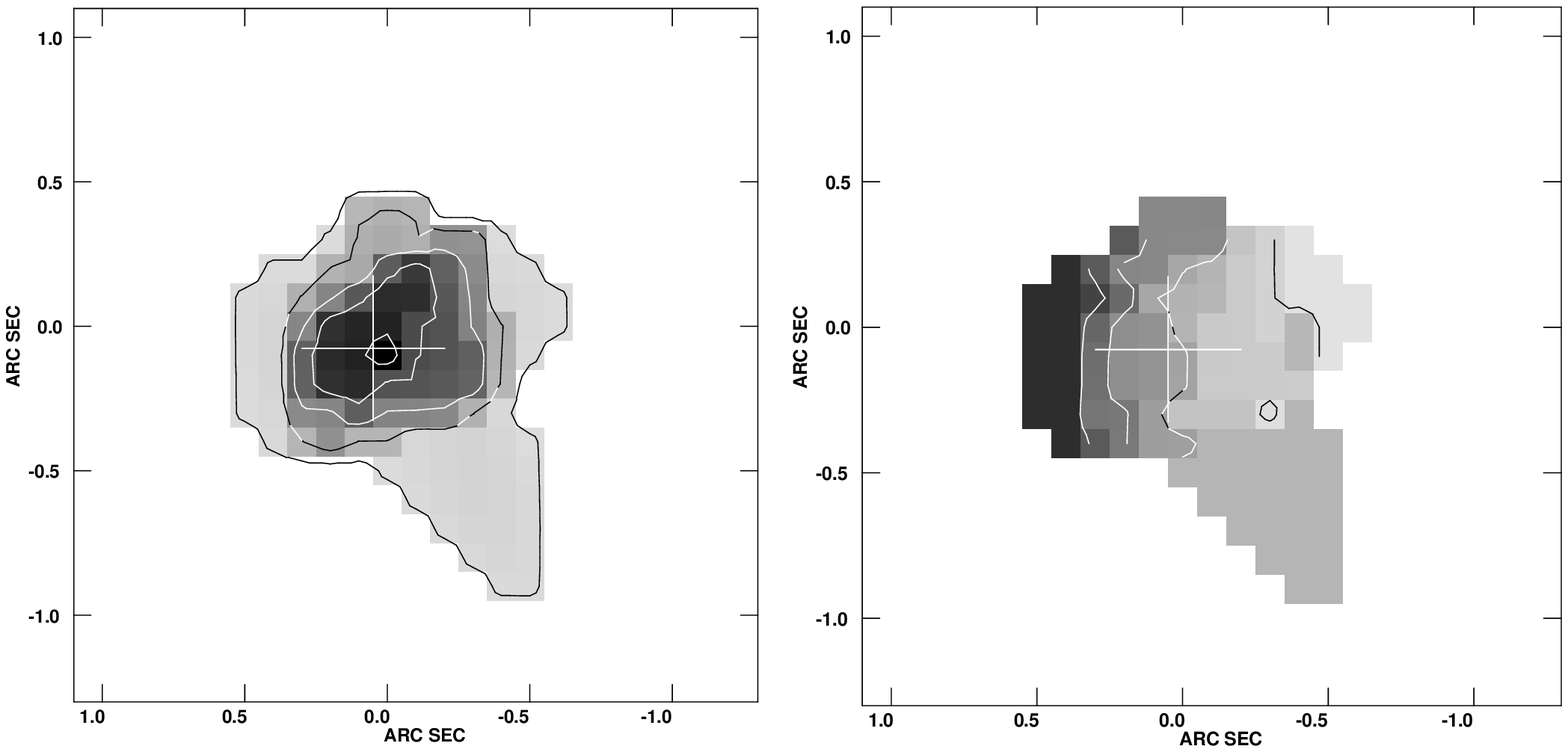}}
\caption{\label{f:disk} Integrated intensity and velocity maps of the high resolution data where only emission in excess of
190~mJy was included to show the location and orientation of the nuclear disk. Position of cross is the same as in Fig.~\ref{f:mom}.
{\it Left panel:}  Integrated intensity where contour levels are $0.72 \times (1,3,5,7,9)$ Jy~beam$^{-1}$~\kms\ and gray scale ranges from
0 to 7.2 Jy~beam$^{-1}$~\kms. {\it Right panel:}  Velocity field with contour levels starting at 1725~\kms\ and then increasing
by steps of 5~\kms. The grayscale ranges from 1720 to 1750~\kms. The synthesized beam is $0.''65 \times 0.''52$. }
\end{figure}


\section{Discussion}
\label{s:discussion}

\subsection{A molecular outflow?}
\label{s:dyn}

We propose that the molecular emission of NGC~1377 is dominated by a disk-outflow morphology.
{\it The notion of a molecular outflow is supported by the presence of  \twco\ 2--1 line wings (Fig.~\ref{f:spec_tco}) and the spatial
orientation of the wing-emission with respect to the line-center emission (Fig.~\ref{f:blue-red}).  The X-shape of the
integrated intensity and dispersion maps (Figs.~\ref{f:mom} and~\ref{f:overlay})  is consistent with a biconical outflow structure.}
In addition, \citet{roussel03} find NIR H$_2$(1â0) S(1) line emission towards the inner 200~pc of NGC~1377 with an orientation
of the velocity field  similar to the one we find for the \twco\ 2--1 line wings. 
Other alternatives to explain the kinematical structure of the \twco\ 2--1 emission include a polar disk or inflowing gas. A
polar disk is unlikely since this would require its velocity to be higher than the rotational velocity of the major axis disk.
That the kinematics is the signature of an outflow, instead of inflowing gas, is supported by the optical dust absorption feature south of the nucleus visible in
Fig.~1 in \citet{roussel06}.  The feature is  spatially coincident with the blue-shifted line wing  suggesting
that the gas on the near side is coming towards us in an outflow. 
Note that non-circular motions in the galaxy plane may be present and will then add to the complexity of the dynamics.

Below we will discuss the morphology and estimates of  molecular mass and velocity of the suggested outflow, but we begin
with a brief discussion of the nuclear disk. The fitted properties of the molecular disk-outflow system are summarized in Tab.~\ref{t:prop}.

\subsubsection{Disk morphology and nuclear dynamical mass}
\label{s:dyn_mass}

\noindent
{\it Disk morphology:} \,To further investigate the notion of a disk-outflow system, we made moment maps of only the brightest emission
(above 190 mJy~beam$^{-1}$)  isolating the narrow line emission emerging from the inner rotating disk (Fig.~\ref{f:disk}).
Note that the fitted PA of the disk of 110$^{\circ}$--130$^{\circ}$ is consistent with  it being perpendicular to the wing emission, the outflow,
while the velocity field (right panel)  is closer to a PA of $\approx$95$^{\circ}$.
This is expected if an outflow is affecting spectral shapes and velocities even close to the nucleus of NGC~1377.
The fitted disk diameter ($D$(FWHM)=$0.\arcsec 60$) should be viewed as a lower limit since we use only the top 23\% of the emission.
Disk inclination is roughly determined to be 45$^{\circ}$ -- 70$^{\circ}$. Uncertainties largely depend on the difficulty in small-scale separation between
disk and outflow components. The stellar large scale disk is estimated to have an inclination
of 60$^{\circ}$ (from the K-band image)  and since it is consistent with the range of values for fits to the nuclear disk we will adopt this value for the nuclear disk as well.\\

\noindent
{\it Disk dynamics:} \, The fitted maximum projected rotational velocity of the disk is 65~\kms\ and for an inclination of 60$^{\circ}$ this
results in a $V_{rot}$ of 75~\kms. 
The dynamical mass of the 60~pc diameter nuclear disk is estimated to M$_{\rm dyn}$ = $4 \times 10^7$ \msun\
using the Keplerian formula M$_{\rm dyn}=2.3 \times 10^8 \times (V_{\rm rot}/100 )^2 \times (r/100) $ (where $V_{\rm rot}$ is in \kms\ and $r$ in pc).
This is only 2.3 times that of the estimated nuclear molecular mass implying a relatively high molecular gas mass fraction in the center of   $\approx$40\%.
A limit to the diameter of the molecular disk (which would be the extension of the nuclear disk) can be set to 200~pc  based on
the total extent of the SMA high resolution \twco\ 2--1  emission along the disk PA.  This is of course sensitivity limited and a fainter, larger molecular extension to
the disk is possible. The Keplerian dynamical mass of this 200~pc disk would be $1.3 \times 10^8$ \msun.

\subsubsection{Outflow morphology}
\label{s:morphology}

\begin{figure}
\resizebox{7cm}{!}{\includegraphics[angle=0]{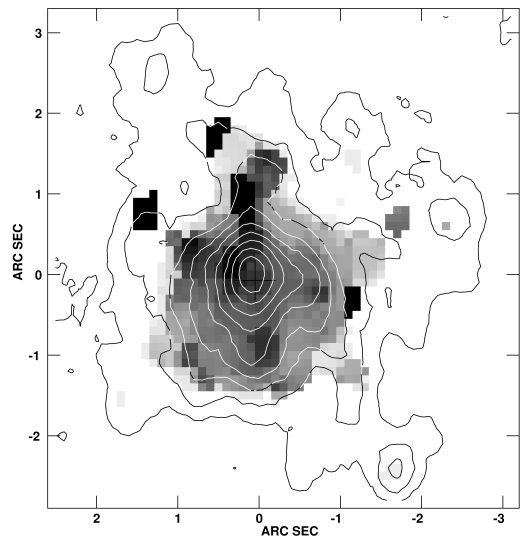}}
\caption{\label{f:overlay} Overlay of high resolution grayscale dispersion map (grayscale ranging from 0 to 34 \kms) on contours of
integrated intensity with contours $2 \times (1,2,3,4,5,6,7,8,9)$ Jy~beam$^{-1}$~\kms. Note how regions of high dispersion form an X-shaped structure generally aligned with a similar structure in the integrated intensity map.
}
\end{figure}

The X-shaped morphology of the high resolution integrated intensity and dispersion maps (see Fig.~\ref{f:overlay}) suggests the molecular gas outlines
a symmetric biconical structure centered on the nucleus.
The cone has an opening angle of 60$^{\circ}$ -- 70$^{\circ}$. Such a large opening angle
could explain the complex velocity pattern observed in the high resolution map. The peak intensity and peak dispersion are roughly correlated in space.
The large dispersion may be caused by line-of-sight effects in maximum path-lengths along the cone walls, and also by multiple spectral components where extended
disk emission overlaps with emission from the outflowing gas.

Note, however, that the molecular outflow may not be well-collimated or in a simple well-ordered outflow pattern, but instead filamentary and patchy and the molecular emission is likely a mix of clumpy and extended gas.  (Properties of wind-cone structures have been modeled by \citet[e.g.][]{tenorio98}). 
The X-shape structure also appears to be evident in the dust morphology of NGC~1377. The optical dust absorption feature south of the nucleus 
(Fig.~1 in \citet{roussel06})  shows a V-shape which is spatially coincident with the blue part of the molecular outflow. The red-shifted molecular outflow is
then hidden behind the galaxy with the blue-shifted outflow projected in front of the disk on the southern side.

\subsubsection{\twco\ luminosity and molecular mass of the outflow}
\label{s:mass_wing}

Without a detailed model  the \twco\ 2--1 luminosity in the outflow is difficult to determine. Below we adopt two approaches to obtain upper and lower limits
to the luminosity.

\noindent
When estimating a {\it lower limit} to the luminosity in the outflow we restrict ourselves to the
(1628 -- 1655 \kms) and (1813 --1840 \kms) part of the plateau component (i.e. 79 \kms\ $\lapprox  |V- V_{\rm sys}|  \lapprox $106 \kms) to ensure that the
rotational components are not included. We find a flux of $13\pm0.3$ Jy~beam$^{-1}$.
Since we are only considering a narrow (27~\kms) part of the 120~\kms\ outflow this will be an underestimate of the
total \twco\ 2--1 flux in the outflow. If the opening angle is as large as we discuss in Sect.~\ref{s:morphology} there will be significant projected outflow emission also at systemic velocities. \\

\noindent
To obtain an {\it upper limit} of the outflow luminosity we fit a Gaussian to the line wings in the low resolution spectrum (see Fig.\ref{f:spec_tco}).
We first fit a Gaussian to the \thco\ 2--1 spectrum (assuming that the \thco\ has only a small contribution from the wings) where
all parameters are free. We obtain: $S_{\rm peak}$=0.15$\pm$0.1 Jy~beam$^{-1}$; $V_{\rm c}$=1740$\pm$2 \kms\ and $\delta V$(FWHM)=61$\pm$5 \kms.
Then, for the \twco\ 2--1, we fit two Gaussian curves, using the parameters from the \thco\ fitting.
The narrow component has $S_{\rm peak}$=1.6$\pm$0.05 Jy~beam$^{-1}$ with $V_{\rm c}$ and $\delta V$ fixed. The broad component
in the wings has, with $V_{\rm c}$ fixed,  $S_{\rm peak}$=0.4$\pm$0.05 Jy~beam$^{-1}$ and $\delta V$(FWHM)=151$\pm$8 \kms\ (and the full
width of the line wings is $\pm 120$ \kms\ (projected)). The division of flux between the line core and outflow component is 104 and
65 Jy~beam$^{-1}$ \kms respectively. Since we detect 20 Jy~beam$^{-1}$ \kms\ in the central high resolution beam, the narrow component flux is mostly located 
outside of the inner 0.\arcsec 6. This flux may be distributed as an extension to the nuclear disk we discuss above in Sec.~\ref{s:dyn_mass}.   

\noindent
To estimate the molecular mass in the outflow we adopt a standard \twco-luminosty to M(H$_2$) conversion factor (see footnote to Tab.~\ref{t:flux}).
This results in M(H$_2$)$_{\rm out}$=$1.1 \times 10^7$ \msun\ for the lower limit and for the upper limit
M(H$_2$)$_{\rm out}$=$5.4 \times 10^7$ \msun. 
The validity of a standard \twco\ to M(H$_2$) conversion factor in galaxies is heavily debated \citep[e.g.][]{narayanan12,wada05,paglione01}.
The global \twco/\thco\ 2--1 line intensity ratio of 9--10 suggests that the molecular gas in NGC~1377 is not optically thin (but \thco\ limits in the line wings are not good
enough for a reliable line ratio for the highest velocities in the outflow). It is also unclear whether the interferometer would pick up emission that is optically thin since it tends to have low surface brightness.  We therefore suggest that applying the conversion factor here does not overestimate gas masses by more
than factors of a few.

\subsubsection{Mass outflow rate, velocities and acceleration.}
\label{s:vel}

The projected peak outflow velocity is $\sim$120~\kms\ (section~\ref{s:kinematics})  and if the opening angle of the outflowing gas is $\sim$$60^{\circ}$
(and the disk inclination is $i$=$60^{\circ}$) then the actual maximum outflow velocity is $\sim$140~\kms.  The outflow has reached
out to $\sim$200~pc (corrected for opening angle and inclination) which, for $V_{\rm out}$=140 \kms\ gives
$t_{\rm dyn}=R/V$=$1.4 \times 10^6$ yrs. The molecular mass in the outflow is $M_{\rm out}=1.1 - 5.4 \times 10^7$ \msun\
(see Sect.~\ref{s:mass_wing}). Thus, the molecular mass loss rate $\delta M/\delta t=\dot{M}$= 8 -- 38 \msun\ yr$^{-1}$. 

We do not know the escape velocity of NGC~1377 hence it is not possible to know whether the gas will leave the system as a whole, or
just the nuclear region of NGC~1377. \citet{martin05} sets a limit for the escape velocity as $V_{\rm esc}$=3$V_{\rm rot}$ and if the rotational 
velocity is 75~\kms\ then the gas will not escape NGC~1377.

An inspection of the central spectrum reveals that broad features are present in the very centre as well suggesting that the gas
is accelerated close to the nucleus (see lower right panel of Fig.~\ref{f:blue-red}). Compared to outflows in M~82 and NGC~253 it appears that the
acceleration to the final velocity happens on a short length scale - smaller than 30~pc.


\begin{table}
\caption{\label{t:prop} Summary of outflow and disk properties$^a$.}
\begin{tabular}{ll}
\hline
\hline \\ 
{\bf Outflow} \\
Molecular mass: & $1.1  - 5.4 \times 10^7$ \msun \\
Extent: & 200~pc \\
Orientation: & PA=35$^{\circ}$ -- 45$^{\circ}$ \\
Opening angle: & 60$^{\circ}$ -- 70$^{\circ}$ \\
Age: & $1.4 \times 10^6$ yrs. \\
Outflow velocity: & 140~\kms \\
$\dot{M}$: & 8 -- 38 \msun\ yr$^{-1}$.\\
Energy: & $2  - 10 \times 10^{54}$ ergs \\
Power: & $0.4 - 2 \times 10^{41}$ ergs~s$^{-1}$  \\
\\
{\bf Nuclear disk}\\
Diameter:  & $0.\arcsec 60$ (60~pc) \\
Orientation: & PA=110$^{\circ}$--130$^{\circ}$ \\
Rotational velocity: & 75 \kms \\
Dynamical mass:  & $4 \times 10^7$ \msun \\
Molecular mass: & $1.7 \times 10^7$  \msun \\
H$_2$ column density:  & $3.4 \times 10^{23}$ $\cmmt$ \\
H$_2$ surface density:  & $5 \times 10^3$ \msun\ pc$^{-2}$ \\

\hline \\
\end{tabular} 

{\it a)} Properties are derived and discussed in  sections \ref{s:kinematics} ; \ref{s:dyn_mass} ; \ref{s:morphology} ; \ref{s:mass_wing} ; \ref{s:vel}

\end{table}


\subsection{Energy and power of the outflow}
\label{s:power}

We use the prescription by \citet{veilleux01} to calculate the kinetic {\it energy} in the outflow: $E_{\rm kin}=E_{\rm bulk} + E_{\rm turbulent}$.
This kinetic energy represents a lower limit to the energy. The work of lifting the gas out of the nucleus
should be included but since we do not know the actual gravitational potential depth of the galaxy this term cannot be added at this stage.
If we assume that the molecular gas in the outflow is moving at constant speed $V_{\rm out}$=140 \kms\ we obtain
$E_{\rm bulk}$=$\Sigma_i (1/2) \delta m_i {v_i}^2$=$(1/2) M_{\rm out} \times  V_{\rm out}^2 $$\approx$$2 \times 10^{54}$ ergs - when
adopting the lower limit to $M_{\rm out}$.  We can however only measure an upper limit to the turbulence of the gas of $\sigma_{\rm 1d}$=25 \kms
(due to effects of spatial resolution, rotation and the outflow) and since this term is already smaller than the contribution from the
bulk motion we do not include it and in total we estimate $E_{\rm kin}$  to $\sim$$2 \times 10^{54}$ ergs. For the upper limit to
$M_{\rm out}$ we get $E_{\rm tot}$$\approx$$10^{55}$ ergs. The power in the outflow is $0.4 - 2 \times 10^{41}$ ergs~s$^{-1}$ which
corresponds to a luminosity of $1 - 5 \times 10^7$ \lsun. The ratio between the wind luminosity and the FIR luminosity is then $2 \times 10^{-3} -  10^{-2}$.
From now on we adopt the more conservative lower value of the luminosity and mass outflow rate. 

\subsection{What is driving the outflow?}
\label{s:driving}

The unusual and obscured nature of NGC~1377 makes it difficult to find an obvious source
of the molecular outflow.  Is it a buried AGN or a young starburst? Below we investigate some possible scenarios. 

\subsubsection{Winds from supernovae and massive stars}
\label{s:winds}

The limit on $L$(1.4 GHz) is $5.3 \times 10^{19}$ W~Hz$^{-1}$ \citep{roussel03}. This corresponds to an upper limit to the SNR of
$5 \times 10^{-4}$ yr$^{-1}$  \citep{condon}. Inserting this limit into equations 10 and 34 in \citet{murray05} we find that SNR fall short
by an order of magnitude to drive the outflow - by momentum or energy. The lack of radio continuum from SNR is consistent
with the nondetection of NIR [Fe II] emission by  \citet{roussel06}.  However, an outflow may also be driven by the wind-momentum
of massive stars. The total momentum in the NGC~1377 outflow is $1.5 \times 10^9$ \msun\ \kms.
Using {\it Starburst99} \citep{leitherer99} we find that a $10^6$ \msun\ cluster
with SFR  of 1/3 \msun\ yr$^{-1}$ has a wind momentum of $10^7$ \msun \kms\ when integrated over 3~Myrs. Thus an SFR of $\sim$30 \msun\ yr$^{-1}$ 
would be required.  From the FIR luminosity we estimate a SFR of 0.4 \msun\ yr$^{-1}$ for NGC~1377 (assuming that all of the FIR is due to star formation) 
and \citet{roussel03} estimate a rate of 1.8 \msun\ yr$^{-1}$. Thus it appears that the SFR is too low by about one order of magnitude to push
out the molecular gas by the wind momentum of massive stars. 

Could the outflow be driven by internal ram pressure from hot gas?  Only upper limits to the thermal free-free emission in the radio has been obtained. 
In Fig.~8 of \citet{roussel03} predictions of the radio
synchrotron and free-free emission based on the IR are plotted. For NGC~1377 the VLA 1.4~GHz upper limit is about a factor of 4-5 below the expected
value.  There is thus no direct evidence of much hot ionized gas. Both the H$\alpha$ and [N II] optical emission lines are weak and
Br$\gamma$ and Pa$\alpha$ are undetected \citep{roussel03,roussel06}.
This may be an effect of the high extinction so to complete the picture of the hot gas of NGC~1377,  X-ray observations will be essential. 

Note also that the symmetry of the wind is inconsistent with the idea of external ram pressure driving the wind, as is
the presence of NGC~1377 in a low density galaxy group where ram pressure forces are expected to be low.

\subsubsection{Radiation pressure}
\label{s:rad_pressure}

\citet{murray05} suggest that radiation pressure (from both star formation and AGNs) from the continuum absorption and scattering of photons on dust grains 
may be an efficient mechanism for driving cold, dusty gas out of a galaxy. 
The lack of evidence for supernovae and the presence of a dusty, compact IR source makes this an attractive scenario for NGC~1377.
\citet{murray05} suggest that for a radiative pressure driven outflow the momentum flux $\dot{M}V$ should be comparable to that in the radiation field
 $L/c$. For NGC~1377 the outflow momentum flux exceeds $L/c$ by factors of a few which is within the uncertainties for outflow structure, velocity
and mass.  We suggest that, with current information, radiation pressure is a potential driving mechanism for the outflowing gas in NGC~1377, but
the fit is far from perfect and other processes cannot be excluded.

\subsubsection{ AGN or starburst?} 
\label{s:agn_sb}

We use the \citet{graham11} (their Figs. 2 - 4 and Tab.~2) calibration of the stellar velocity dispersion and BH mass to estimate the mass of a 
supermassive black hole (SMBH)  in
NGC~1377. \citet{wegner03}  list a value of the stellar velocity dispersion of $\sigma$=83 \kms\ for NGC~1377 in their online database.
This gives a SMBH mass in the range of about $4 \times 10^5 - 6 \times 10^6$ \msun\ with a most likely value of about $1.5 \times 10^6$ \msun.
If all the FIR emission observed would be due to accretion onto the SMBH it would then operate at $\sim$10\% of its Eddington limit.
A small acceleration region is consistent with a compact source of energy and momentum.
Radiation pressure acceleration minimizes hard shocks which results in low radio emission and acts on dense gas, consistent with the
IR H$_2$ emission and the suggested \twco\ 2--1 molecular outflow. It is also easier to bury a compact energy source behind the gas and dust, giving
rise to the deep silicate absorption. Strong FIR emission lines from ionized gas in an AGN are expected to be highly localized around the SMBH,
and thus are easier to hide.

In contrast,  a dust-enshrouded nuclear starburst needs to be located well inside a radius of 29 pc (to launch the outflow). The current molecular mass here is
estimated to $1.7 \times 10^7$ \msun, 40\% of the dynamical mass. \citet{roussel06} suggest a starburst with a mass $1.6 \times 10^7$ \msun\ assuming an age of 1~Myr, a Salpeter IMF between 0.1 and 120 \msun\  and $3 \times 10^4$ O-stars to produce the bolometric luminosity of NGC~1377. The star formation efficiency (SFE) 
of the starburst then had to be at least 50\% and all of its Pa$\alpha$, and Br$\gamma$ emission has to be absorbed. This implies that the stars must be deeply buried inside the nuclear dust cloud putting even stricter limits on their radial distribution.  In addition, the starburst scenario requires that almost all of the stellar mass in the
center of NGC~1377 has to be very young.


\subsection{Comparing with other molecular outflows}
\label{s:comp}

Outflowing molecular gas has been detected towards several galaxies of various luminosities. In some cases the
existence of an outflow has been inferred from \twco\ line wings - such as for the LIRG NGC~3256 \citep{sakamoto06}.
In some cases, the molecular outflow was discovered in already iconic optical outflows, such as is the case for M~82
\citep{nakai87,walter02}. There is also mounting evidence of massive, $V\gapprox1000$ \kms\  outflows in AGN/starburst driven
ULIRGs \citep[e.g.][]{sakamoto09,feruglio10,chung11,sturm11,aalto12}.

NGC~1377 can also be compared to the AGN-driven molecular outflow of the S0 galaxy NGC~1266 \citep{alatalo11}.  Nuclear gas depletion
time scales are similar to those estimated for NGC~1377, and both galaxies have compact, $r<100$~pc nuclear disks confining
the outflowing wind. NGC~1266 is more FIR luminous than NGC~1377 with an order of magnitude more molecular gas and its
molecular disk has a factor of five greater gas surface density.  Also, NGC~1266 is not radio deficient ($q$=2.3) unlike NGC~1377.
Estimating the power of a radio jet from the total radio flux, \citet{alatalo11} attributed the molecular outflow to the mechanical work
of the radio jet of the AGN (see also \citet{matsushita07} for a similar case). 

What sets NGC~1377 apart is its unusual IR properties and extreme $q$ suggesting that the nuclear activity is in a transient phase of its evolution
or that NGC~1377 is exceptional in some other aspect. This is illustrated by the fact that observationally the $q$ parameter is in a strikingly
small range in large statistical samples of galaxies. In particular, the upper-bound of q is very clear and $q>3$ is really exceptional
(see for example Fig.~6 of \citet{yun01}.

Comparing the energy and mechanical luminosity of the NGC~1377 outflow to one AGN-driven (NGC~1266) and one starburst driven
(NGC~3256) outflow we find them to be similar (within factor of a few). The ratio between the wind luminosity and the FIR
luminosity $L_{\rm mech}/L_{\rm FIR}$, however, for NGC~1377 is similar to, or higher, than that of NGC~1266,
and at least one order of magnitude higher than that of the starburst driven outflow of NGC~3256. More studies will show if
this difference is related to the underlying driving mechanism. The $L_{\rm mech}$ is calculated only on the molecular flows in all three galaxies.

\subsection{Nuclear gas properties}
\label{s:nuclear}

\subsubsection{Is there a Compton-thick nuclear dust and gas cloud?}
\label{s:compton}

For NGC~1377 to hide its power source also at X-ray wavelengths large columns of absorbing material
has to reside in front of the nucleus. We estimate the average $N$(H$_2$) in the inner $r$=29~pc of NGC~1377 to
be $3.2 \times 10^{23}$ $\cmmt$. (This corresponds to an $A_{\rm V}$ of 290 (for
$N_{\rm H}$ ($\cmmt$) = $(2.21 \pm 0.09) \times 10^{21} A_{\rm V}$ \citep{guver09}.)
For an absorber to be Compton-thick an $N_{\rm H}$$>$$10^{24}$ $\cmmt$ ($A_{\rm V}$$>$450) is required.
Thus, a Compton-thick absorber, if any, should therefore be smaller than our beam. 
If the \twco\ luminosity of our \twco\ 2--1 beam is located in a region with a radius $\lapprox$17~pc the
resulting $N_{\rm H}$ would be $\gapprox 10^{24}$ $\cmmt$.

\subsubsection{Gas physical conditions}
\label{s:excitation}

We find a peak \twco\ 2--1 temperature of 19~K in a 58~pc beam which serves as a strict lower limit to the gas kinetic temperature
in this region. The dust temperature is difficult to determine but a fit to the 25--100 $\mu$m data results in a black~body temperature
of 80~K with a diameter of 37~pc \citep{roussel03}. Comparing our 1~mm continuum flux to the FIR SED suggests it is consistent with a $\nu^3$ slope - either implying
$\beta$=1 - or that the dust is already opaque at $\lambda$=100 $\mu$m. This possibility, if confirmed,
would be consistent with the dusty opaque core suggested from the deep silicate absorption.
There is no evidence of an excess of 1~mm flux emerging from cold (10-20 K) dust.

\noindent
We can consider three simple scenarios for the physical conditions of the nuclear molecular gas:

\begin{enumerate}

\item {\it The molecular gas is cold and optically thick.} The \twco\ 2--1 emission is optically thick, fills the central beam and its brightness temperature
traces the kinetic temperature of the gas at 29~pc radius from the nucleus. This requires a very steep, radial temperature gradient
to allow for the buried warm dust.

\item {\it The molecular gas is optically thin} In this scenario, the \twco\ 2--1  brightness temperature does not reflect
the kinetic temperature of the gas even if it fills the central beam and will instead be $\propto T_{\rm ex} \tau$. Given the
large columns of absorbing gas and dust this scenario seems unlikely.  

\item {\it The molecular gas is warm and optically thick}  A molecular medium of warm gas clumps that only fills
a fraction of the beam. Alternatively, the \twco\ peak is unresolved in the beam. If we assume that the emission is
emerging from the same region as the inferred dust diameter (of 37~pc) the peak \twco\ surface brightness temperature would be 50~K.

\end{enumerate}

The above scenarios assume a single-component ensemble of clouds. A more realistic view on the nuclear molecular ISM is that it likely
consists of both dense ($n>10^4$ $\cmmd$) gas clumps embedded in a lower density ($n=10^2 - 10^3$ $\cmmd$) molecular medium
(see discussion in \citep[e.g.][]{aalto94}).
The \twco\ 2--1 brightness will then reflect the relative filling factor of the low- and high density gas.  The detection of bright HCN 1--0 emission
\citep{imanishi09} towards the inner 2\arcsec of NGC~1377 suggests gas densities $>$$10^4$ $\cmmd$ (unless the HCN is IR pumped). 

Is the molecular gas concentration forming stars? From the Kennicutt-Schmidt (KS) relation \citep{kennicutt98}  the expected SFR from the
nuclear gas concentration is 0.1\msun\ yr$^{-1}$. This falls short by factors of 4 -- 18 to explain the inferred SFR if all FIR emission is emerging
from star formation. However, the KS relation at $r$$<$100~pc may should have a large scatter since star formation is unlikely to be a steady-state process
on such small scales. \citep[e.g.][]{onodera10}. There should be some star formation going on even if it is difficult to infer the actual SFR.

\subsection{The evolutionary implications for NGC~1377}
\label{s:evolution}

\subsubsection{Origin of the molecular gas}

Just as is the case for NGC~1266 \citep{alatalo11}, one question for NGC~1377 is how the gas got to the very center
of the galaxy so efficiently. So far, no clear evidence of an interaction has been presented.
The range of possible PAs of the nuclear disk includes a difference by about 40$^{\circ}$compared to the stellar disk
of NGC~1377. An inclination difference between the gas and stellar disks would indeed suggest an external source for
the gas \citep{davis11}. It is interesting that the evidence so far suggests that 
most of the gas has ended up in molecular form. If NGC~1377 accreted a late type dwarf galaxy or gas from
the outer regions of a large disk galaxy, the gas is more likely to be atomic.  Somehow the gas must have become shocked and compressed along
the way. However, the lack of 1.4~GHz radio continuum emission makes it difficult to assess the amount of cold
atomic hydrogen in the center of NGC~1377 - which normally could be tested by searching for absorption in the 21~cm HI line.

The gas concentration may instead be the result of slow, secular evolution such as accretion along a bar \citep[e.g.][]{jogee05}, very slow external
accretion (galaxy harassment), or stellar return \citep[e.g.][]{welch03}. All of these - or combinations of them - are possible scenarios, but the problem
remains of getting such a concentration of molecular gas to the core of NGC~1377.

\subsubsection{The nuclear activity}

Observing the molecular properties of NGC~1377 allow us to probe the early stages of nuclear activity and feedback mechanisms
in active galaxies. Depending on what is driving the FIR emission and outflow of NGC~1377 we are either 
witnessing the growth and feeding of a black hole in an early type galaxy - a process fueled by the late  influx 
of molecular gas. Alternatively,  the gas fuels the early triggering phase of a nuclear starburst increasing the stellar mass instead of the
central black hole. 

Another important key to the evolutionary stage and nature of the activity is the age of the outflow. If the SMA data recovers the full extent of the
outflow, then this sets an upper limit to its age of 1.4~Myrs. High sensitivity observations of \twco\ 1--0 as well as deep imaging of dust absorption
features, H$\alpha$ and a search for HI will help determine if the outflow extends beyond the \twco\ 2--1 emission.  Such evidence would
further constrain the age of a buried power source. An implicit assumption here is that this activity is a one-time event (i.e., not recurrent in a short time scale).
The ongoing, mass outflow rate would clear the central region of gas within the short time period of 5 -- 25 Myrs.  The growth of the SMBH
and/or nuclear star formation will cease within this time period.  However, if the gas is unable to escape the galaxy we might be observing
a failed wind \citep[e.g.][]{dorod11} . The gas may then fall back onto the galaxy again where the returned gas
serves as a reservoir for future activity. Further studies will address this possible evolutionary scenario.

\section{Conclusions}

We have imaged \twco\ and \thco\ 2--1 in the FIR-excess galaxy NGC~1377 with the SMA:

\begin{enumerate}
\item We find bright, complex \twco\ 2--1 line emission from the nuclear region of the FIR-excess S0 galaxy NGC~1377
the structure of which is reminiscent of a disk-outflow system. The \twco\ 2--1 line wings, the spatial
orientation of the wing-emission with respect to the line-center emission, its correspondence with the optical dust features and the X-shape of the
integrated intensity and dispersion maps support this interpretation. Estimated outflow parameters are: $M_{\rm out}$(H$_2$)$>$$1 \times 10^7$ \msun;
$\dot{M}$$>$8 \msun\ yr$^{-1}$; $V_{\rm out}$$>$140 \kms; extent$\approx$200~pc; opening angle=$60^{\circ} - 70^{\circ}$; PA$\approx$$40^{\circ}$
and an age of $\sim$1.4$\times 10^6$ yrs.\\

\item We suggest that the age of the outflow supports the notion that the nuclear activity is young - a few Myrs. The dusty center of NGC~1377, the lack of significant
numbers of recent supernovae, or of a dominant hot wind, may imply that the outflow is driven by radiation pressure from a compact, dust enshrouded
nuclear source.  However,  other driving mechanisms are possible.  Accretion onto a $10^6$ \msun\ SMBH -  i.e. a buried AGN - is a possible power source. Alternatively, the photons may emerge from a very young (1~Myr), pre-supernova starburst, but we find this explanation less likely due to the extreme
requirements of compactness and youth of the source.
The implied mass outflow rate is large enough that it will clear the nuclear region of gas within  5--25 Myrs  and will, at least temporarily, shut off the nuclear activity and
growth. If the outflowing gas rains back onto the galaxy, however, it may serve as fuel for a future period of nuclear activity and/or star formation.  \\

\item Molecular masses are estimated through the adoption of a standard \twco\ to H$_2$ conversion factor - with all the implied uncertainties. In the whole \twco\ 2--1 map we detect M(H$_2$)=$1.5 \times 10^8$ \msun\ and in the inner $r$=29~pc M(H$_2$) is estimated to $1.7 \times 10^7$ \msun\  with an  H$_2$ column density of $N$(H$_2$)$= 3.4 \times 10^{23}$ $\cmmt$. This is  not large enough for the core to be Compton thick and a Compton-thick absorber, if any, should therefore be smaller than our beam. The average gas surface density within $r$=29~pc is $\Sigma$=$5 \times 10^3$ \msun\ pc$^{-2}$. \\

\item The \twco\ 2--1 brightness temperature sets a lower limit to the nuclear gas kinetic temperature of 19~K, but if cloud filling factors are less then unity then gas
temperatures will be significantly higher.

\end{enumerate}

\begin{acknowledgements}
      The Submillimeter Array is a joint project between the Smithsonian
Astrophysical Observatory and the Academia Sinica Institute of Astronomy and
Astrophysics and is funded by the Smithsonian Institution and the Academia
Sinica. JSG thanks the U.S. National Science Foundation for partial support of
this work through NSF grant AST-0708967 to the University of
Wisconsin-Madison. SA thanks the Swedish Research Council for support 
(grant 621-2011-4143). KS is supported by the Taiwan NSC grant 99-2112-M-001-011-MY3.
SM acknowledge the co-funding of this work under the Marie Curie Actions
of the European Commission (FP7-COFUND).
\end{acknowledgements}

\bibliographystyle{aa}
\bibliography{n1377_ref}
\end{document}